\documentclass{pasj00}

\begin{document}
\SetRunningHead{K. Motogi et al.}{Accelerating a water maser face-on Jet from a high mass young stellar object}
\Received{2014/10/21}
\Accepted{2014/10/29}
\Published{2015/02/02}

\title{Accelerating a water maser face-on jet from a high mass young stellar object. }

\author{%
  Kazuhito  \textsc{motogi}, \altaffilmark{1} Kazuo  \textsc{sorai}, \altaffilmark{2} Mareki  \textsc{honma}, \altaffilmark{3, 4} Tomoya  \textsc{hirota} \altaffilmark{3, 4}, 
Kazuya \textsc{hachisuka} \altaffilmark{1}, \\Kotaro \textsc{niinuma} \altaffilmark{1, 5},  Koichiro  \textsc{sugiyama} \altaffilmark{6},   Yoshinori  \textsc{yonekura} \altaffilmark{6}, and Kenta  \textsc{fujisawa} \altaffilmark{1, 5}}
\altaffiltext{1}{The Research Institute of Time Studies, Yamaguchi University, Yoshida 1677-1, Yamaguchi-city, Yamaguchi 753-8511,Japan}
\altaffiltext{2}{Department of Physics / Department of Cosmosciences, Hokkaido University, Kita 10, Nishi 8, Kita-ku, Sapporo, Hokkaido 060-0810, Japan}
\altaffiltext{3}{Mizusawa VLBI Observatory, NAOJ, 2-21-1 Osawa, Mitaka, Tokyo 181-8588, Japan}
\altaffiltext{4}{Department of Astronomical Science, Graduate University for Advanced Studies (SOKENDAI), 2-21-1 Osawa, Mitaka, Tokyo 181-8588, Japan}
\altaffiltext{5}{Department of Physics, Faculty of Science, Yamaguchi University, Yoshida 1677-1, Yamaguchi-city, Yamaguchi 753-8512, Japan}
\altaffiltext{6}{Center for Astronomy, Ibaraki University, 2-1-1 Bunkyo, Mito, Ibaraki 310-8512, Japan}

\email{motogi@yamguchi-u.ac.jp}

\KeyWords{stars:formation -- ISM:jets and outflows -- masers} 

\maketitle

\begin{abstract}
We report on a long-term single-dish and VLBI monitoring for intermittent flare activities of a Dominant Blue-Shifted H$_{2}$O Maser (DBSM) associated with a southern high mass young stellar object, G353.273+0.641. 
Bi-weekly single-dish monitoring using Hokkaido University Tomakomai 11-m radio telescope has shown that a systematic acceleration continues over four years beyond a lifetime of individual maser features. 
This fact suggests that the H$_{2}$O maser traces a region where molecular gas is steadily accelerated. 
There were five maser flares during five-years monitoring, and maser distributions in four of them were densely monitored by the VLBI Exploration of Radio Astrometry (VERA). 
The overall distribution of the maser features suggests the presence of a bipolar jet, with the 3D kinematics indicating that it is almost face-on (inclination angle of $\sim$ 8$^{\fdg}$--17$^{\fdg}$ from the line-of-sight). 
Most of maser features were recurrently excited within a region of 100$\times$100 AU$^{2}$ around the radio continuum peak, while their spatial distributions significantly varied between each flare. 
This confirms that episodic propagations of outflow shocks recurrently invoke intermittent flare activities. 
We also measured annual parallax, deriving the source distance of 1.70 $^{+0.19}_{-0.16}$ kpc that is consistent with the commonly-used photometric distance. 
\end{abstract}

\section{Introduction}
Astronomical masers such as H$_{2}$O and SiO are unique tools for studying mass-loss activities in high mass star formation. 
These maser species are basically excited in strong shocks on a working surface between a protostellar outflow and a dense envelope (e.g., \cite{Furuya2000}; \cite{Moscadelli2000}). 
The masers allow us to directly observe an intrinsic outflow driven by individual High Mass Young Stellar Objects (HMYSOs) within 10$^{3}$ AU scale, minimizing any contamination from a mass entrainment and/or outflows driven by nearby cluster members. 
Outflow properties such as size, velocity and morphology, predicted from interferometric maser studies vary source to source, e.g., 
collimated jet (\cite{Shepherd2004}), disk wind (\cite{Matthews2010}; \cite{Greenhill2013}), wide angle outflow (\cite{Motogi2008}), expanding shell (\cite{Torrelles2003}; \cite{Kim2013};\cite{Surcis2014}), equatorial outflow (\cite{Trinidad2007}) or combination of jet and wide angle outflow (\cite{Torrelles2011}). 
Such divergences are possibly related to a number of factors such as stellar mass, evolutionary stage, geometry of surrounding envelope and driving mechanism of host outflows. 

\citet{Motogi2011b} (Hereafter Paper I) focused on a highly variable H$_{2}$O maser source associated with a southern HMYSO, G353.273+0.641 (Hereafter G353). 
They have found two intermittent maser flares during two years, combining a frequent monitoring by Very Long Baseline Interferometry (VLBI) and single-dish observation. 
In Paper I, they proposed that such a flare activity is recurrently excited by episodic shock propagations caused by a high-velocity protostellar jet. 
If this is the case, one can extract time-domain information of a jet-launching region from a maser variability, 
since bright maser emission allows us easy and frequent monitoring even with a small size radio telescope. 
Such a region is expected to be so small ($\sim$ a few AU or less, e.g., \cite{Machida2008}; \cite{Seifried2012}), compared to current observational resolutions. 

G353 is embedded in the high mass star-forming region NGC6357. 
\citet{Neckel1978} determined its photometric distance as 1.74 $\pm$ 0.31 kpc from the Sun. 
Exact YSO mass is still unclear, however, association of a class II CH$_{3}$OH maser emission suggests that the source is already a high mass object (\cite{Caswell2008}, hereafter CP08). 
G353 is known as the Dominant Blue-Shifted Maser (DBSM) source that is a special class of a 22 GHz H$_{2}$O maser sources (CP08). 
DBSMs have a highly asymmetric spectrum, where almost all the flux is concentrated in highly blue-shifted components.  
The peak line-of-sight (LOS) velocities are ranged from -60 to -45 km$^{-1}$ against the systemic velocity ($V_{\rm sys}$) of $\sim$ -5.0 km s$^{-1}$ in the case of G353, 
adopted from the peak velocity of the class II CH$_{3}$OH maser emission (CP08). 
CP08 proposed that DBSMs are excited by a well-collimated jet in a nearly face-on geometry. 
\citet{Motogi2013} have actually detected both of a faint radio jet and an extremely high-velocity thermal SiO jet, supporting CP08. 

Paper I also found a signature of maser acceleration along the LOS in two velocity components ($\sim$ -53 km s$^{-1}$ and 73 km s$^{-1}$) accompanied by a significant flare-up. 
Such an acceleration was observed in both of the single-dish and VLBI data, excluding instrumental errors. 
Despite a similar acceleration rate ($\sim$ -5 km s$^{-1}$ yr$^{-1}$), these two components were spatially independent. 
This fact confirmed that observed acceleration was not an apparent phenomenon caused by Christmas-tree effect and/or any variation of internal structures of individual maser features, 
but a real one driven by global momentum supply related to shock propagations. 

In this paper, we report on further monitoring observations of the H$_{2}$O maser at 22.23508 GHz ($J_{K_{\rm a}K_{\rm c}}$ = $6_{16}$--$5_{23}$) towards G353. 
We have studied variability and recurrence in spatial distributions of maser features between each flare, employing denser VLBI monitoring intervals than our previous work. 
New single-dish data validated the systematic acceleration of the masers. 

\section{Observations and data reduction}
\subsection{VLBI monitoring}
The VLBI monitoring was performed by the VLBI Exploration of Radio Astrometry (VERA). 
Typical observing interval was $\sim$ 45 days that is roughly half of that in Paper I. 
We had total 15 epochs between 2010 September to 2012 August. 

All observations were made in the dual beam mode, where target and phase calibrator were observed simultaneously (\cite{Kawaguchi2000}; \cite{Honma2003}). 
Total on-source time was $\sim$ 5 hours in each epoch. 
The phase tracking center for G353 was ($\alpha$,$\delta$)$_{\rm J2000.0}$ = (\timeform{17h26m01.59s}, \timeform{-34D15'14.9"}). 
We used a bright quasar, J1717-3342 ($\sim$ 700 mJy) for the dual-beam phase referencing. 
The separation angle between G353 and J1717-3342 is \timeform{1D.83} with a position angle of -\timeform{108D}. 
NRAO530 and/or PKS1510-089 was observed once per hour for delay and bandpass calibration.  
Antenna gains were corrected based on measured system noise temperatures ($T_{\rm sys}$) of each stations. 
Typical $T_{\rm sys}$ are ranged 200 -- 400 K for Mizusawa and Iriki stations and 400 -- 800 K for Ogasawara and Ishigaki stations. 

Details of backends and correlation processes were already described in Paper I. 
We observed single left-hand circular polarization with a data rate of 1024 Mb s$^{-1}$ (2 bit, 512 MHz sampling). 
The total bandwidth was 240 MHz (16 MHz $\times$ 15 IF) for calibrators and 16 MHz (1 IF) for G353. 
We cut out an 8-MHz sub-IF from the maser IF and divided it into 512 spectral channels in initial two epochs, as it was in Paper I. 
The total velocity coverage is 108 km s$^{-1}$ with spectral resolution of 0.21 km s$^{-1}$. 
On the other hand, we used two 8-MHz sub-IFs in latter 13 epochs, in order to cover a larger velocity range in G353. 
Central frequencies of these sub-IFs are separated 4 MHz, as both including maser emission peaks. 
This gives a velocity coverage of 162 km s$^{-1}$ with the same spectral resolution. 

Data reduction was carried out in the standard reduction procedure for VERA data (see \cite{Motogi2011a}), 
using the Astronomical Imaging Processing System (AIPS) package developed by National Radio Astronomy Observatory (NRAO). 
\footnote{The National Radio Astronomy Observatory is a facility of the National Science Foundation operated under cooperative agreement by Associated Universities, Inc.}
We first determined an astrometric frame in each epoch from phase-referenced positions of bright maser spots with respect to J1717-3342, and then, fringe-fitting and self-calibration were performed using the brightest maser channel in each epoch. 
We finally searched all maser spots from the self-calibrated data cube with a detection limit of 6 $\sigma$. 
Searched area was \timeform{1.5"}$\times$\timeform{1.5"} from the phase tracking center. 
An interval of phase solution was $\sim$2 min for a fringe-fitting using J1717-3342, and down to 15 sec in final self-calibration. 

Only a maser emission that appeared in successive three or more spectral channels in at least one observing epoch was treated as a real detection. 
Each set of spatially successive maser spots was defined as a 'maser feature' that means individual gas clump, showing a single-peaked velocity profile.  
Velocities and positions of each feature were determined by the flux-weighted averaging. 

We included thermal image noises,  baseline solution errors and tropospheric zenith delay residuals as astrometric error sources. 
The third term was estimated within a typical range of $\pm$ 3 cm, by the image-optimization method \citep{Honma2008}. 
This estimation, however, seemed to be incomplete in some epochs, since image dynamic ranges were still limited even after the correction. 
We deduced overall errors of the image optimization, using two distinct sub-IF data which were available in most epochs. 
We analyzed each dataset independently and compared astrometric positions of maser spots detected in both data. 
These positions were basically consistent within an error of thermal image noises, but some epochs show a significant positional offset up to 3 milli-arcsecond (mas). 
To be conservative, we included it in total astrometric errors as a systematic error of our analysis, since frequency difference of only 4 MHz is unlikely to cause such a large offset. 
The relative positional accuracies in the self-calibrated data, which were dominated by thermal image noises, are better than 0.03 and 0.1 mas for the right ascension and declination, respectively. 

Table \ref{tab:obs} summarizes all VERA observations including that reported in Paper I. 
We note that all tables in this paper are presented in Appendix section. 
It contains observing dates, synthesized beams ($\theta_{\rm b}$), beam position angles (PA) east of north, 
typical image noise levels ($\Delta I$), astrometric errors ($\sigma_{1}$), residual systematic errors ($\sigma_{2}$) and any special comments. 
Table \ref{tab:feature} listed only features detected in three or more epochs, which were used for kinematic analyses in the following sections. 
All data of detected maser features are available in supplementary Table \ref{tab:obs}.

\subsection{Single-dish monitoring}
Single-dish monitoring was performed by the Hokkaido University Tomakomai 11-m Telescope \citep{Sorai2008}. 
Follow-up monitoring was started from 2010 November and completed 2013 March. 
Observing intervals were denser than bi-weekly, except for maintenance sessions during summer seasons. 
The beam size and aperture efficiency were 4.2 arcmin and 49 per cent, respectively. 
Detailed procedure of observations and data reductions using the NEWSTAR software package developed at Nobeyama Radio Observatory were already described in Paper I. 
The total velocity coverage and final spectral resolution are 216 and 0.21 km s$^{-1}$, respectively.  
The rms noise levels were $\sim$ 2-3 Jy in winter seasons and increased up to 15 Jy in the worst case in summer seasons. 

\section{Results}
\subsection{Light Curve and Dynamic Spectrum}
\begin{figure}[htb]
\begin{center}
\FigureFile(90mm,90mm){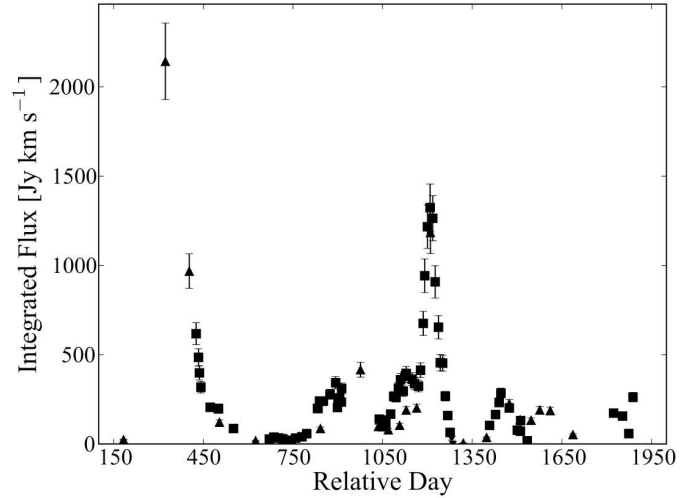}
\end{center}
 \caption{Light curve for the main velocity component. 
 January 1st, 2008 is counted as the first day. 
 Filled squares and triangles correspond to the single-dish and VLBI data. 
 The flux was integrated over the veracity range (-65 to -45 km s$^{-1}$). 
 } \label{fig:light}
\end{figure}

\begin{figure}[htb]
\begin{center}
\FigureFile(90mm,90mm){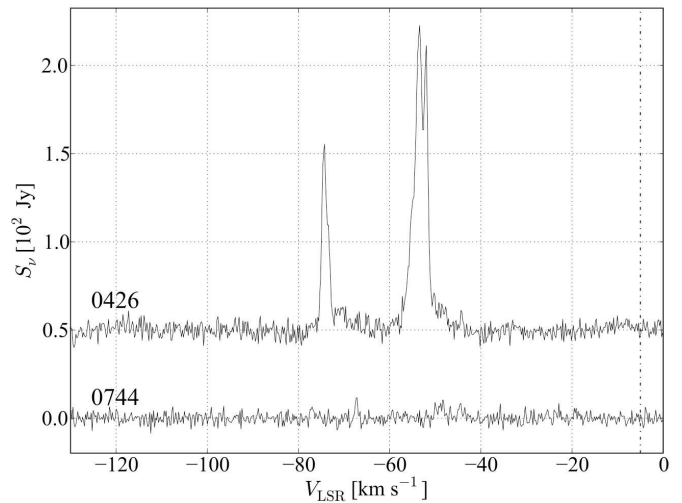}
\end{center}
 \caption{Typical single-dish spectra in both flare-up phase (day $\sim$ 426) and quiescent phase (day $\sim$ 744). 
We added flux offset of 0.5 Jy in the former. 
The dot-dashed line indicates $V_{\rm sys}$ of -5.0 km s$^{-1}$. 
 } \label{fig:spectrum}
\end{figure}

Figure \ref{fig:light} shows the light curve of the main velocity component (-65 -- -45 km s$^{-1}$) that contains almost all maser flux in G353. 
Typical single-dish spectra in both flare-up and quiescent phases were also plotted in Figure \ref{fig:spectrum}. 
We have detected five maser flares during five years, and four except for the last one were also monitored by VERA. 
The durations of these four flares were defined as, 1st: 300 -- 500, 2nd: 800 -- 950, 3rd: 1050 -- 1300, 4th: 1400 -- 1690, using the relative day counted from January 1st, 2008. 
The target showed intermittent flare-up at a typical interval of $\sim$ 1.5 years, although no clear periodicity was seen as reported in Paper I. 
A higher velocity component ($<$ -70 km s$^{-1}$), which showed a significant flare-up in the initial flare, has been staying quiescent state after the second flare. 
We show a spectral movie made by all single-dish data in supplementary Figure 1. 

\begin{figure*}[htb]
\begin{center}
 \FigureFile(134mm,134mm){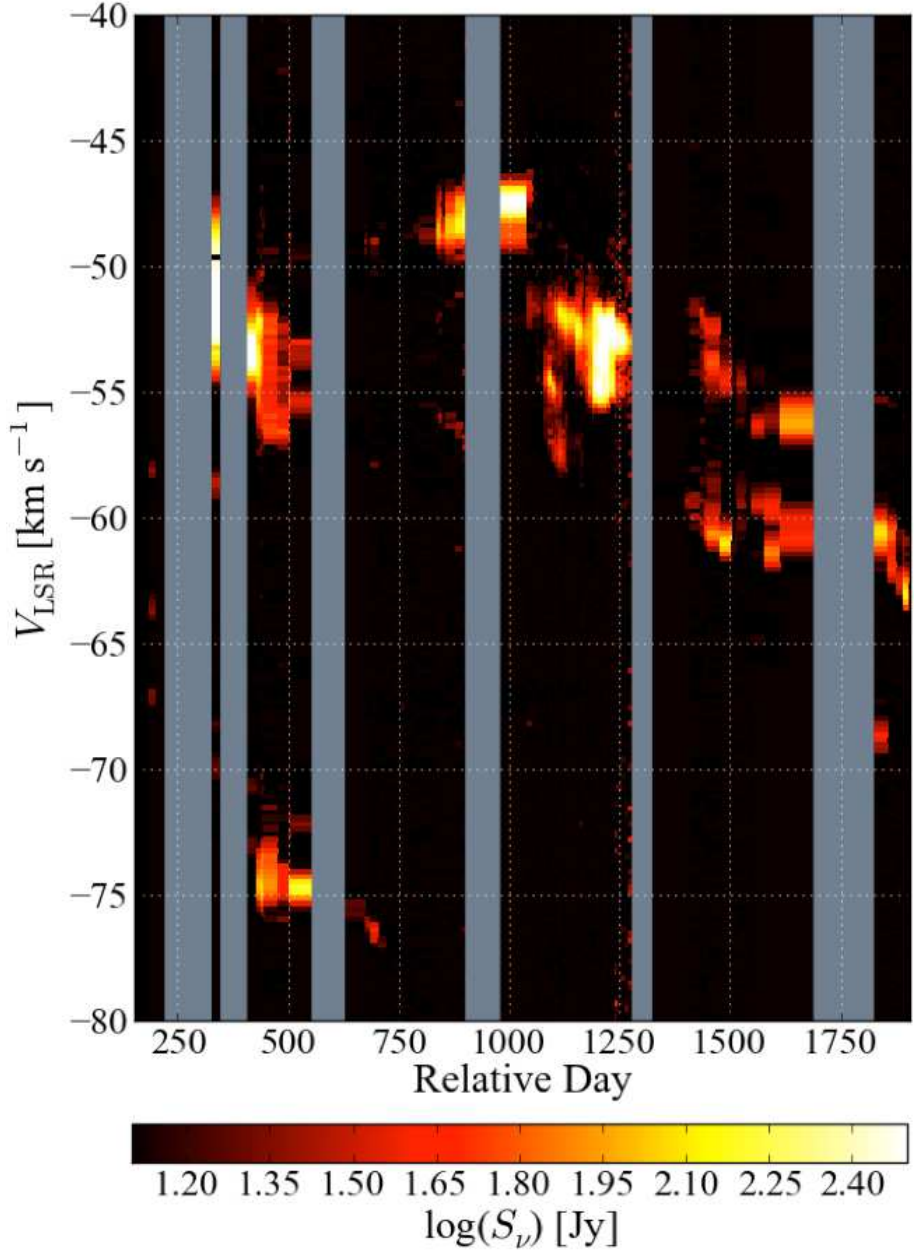}
\end{center}
 \caption{Dynamic spectrum in a veracity range of -40 to -80 km s$^{-1}$. 
 The grey shaded durations correspond maintenance sessions of the Tomakomai 11-m telescope. 
 Color version of this figure is available online.  
 } \label{fig:dynamic}
\end{figure*}

Figure \ref{fig:dynamic} presents a dynamic spectrum with a velocity range of -80 to -40 km s$^{-1}$.  
The spectrum contains both of the single-dish and VLBI data. 
The velocity widths were sometimes narrower in the latter data, because of resolved-out flux (e.g., day $\sim$ 840).  
There is a clear signature of a systematic acceleration in several velocity components. 

As already mentioned, accelerations of two velocity components in the 1st flare-up ($\sim$ -53 and -73 km s$^{-1}$) were already reported in Paper I. 
These spatially independent components have shown a similar acceleration rate ($\sim$ -5 km s$^{-1}$ yr$^{-1}$). 
After the 2nd flare-up, the newly appeared component around -47 km s$^{-1}$ had been continuously accelerated up to -63 km s$^{-1}$ during 3 years. 
The linearly-fitted acceleration rate of this new component is -5.4 $\pm$ 0.18 km s$^{-1}$ yr$^{-1}$, where velocity dispersions were used as a fitting weight. 
This is consistent with that of the two components in the 1st flare-up. 

\begin{figure*}[htb]
\begin{center}
 \FigureFile(150mm,150mm){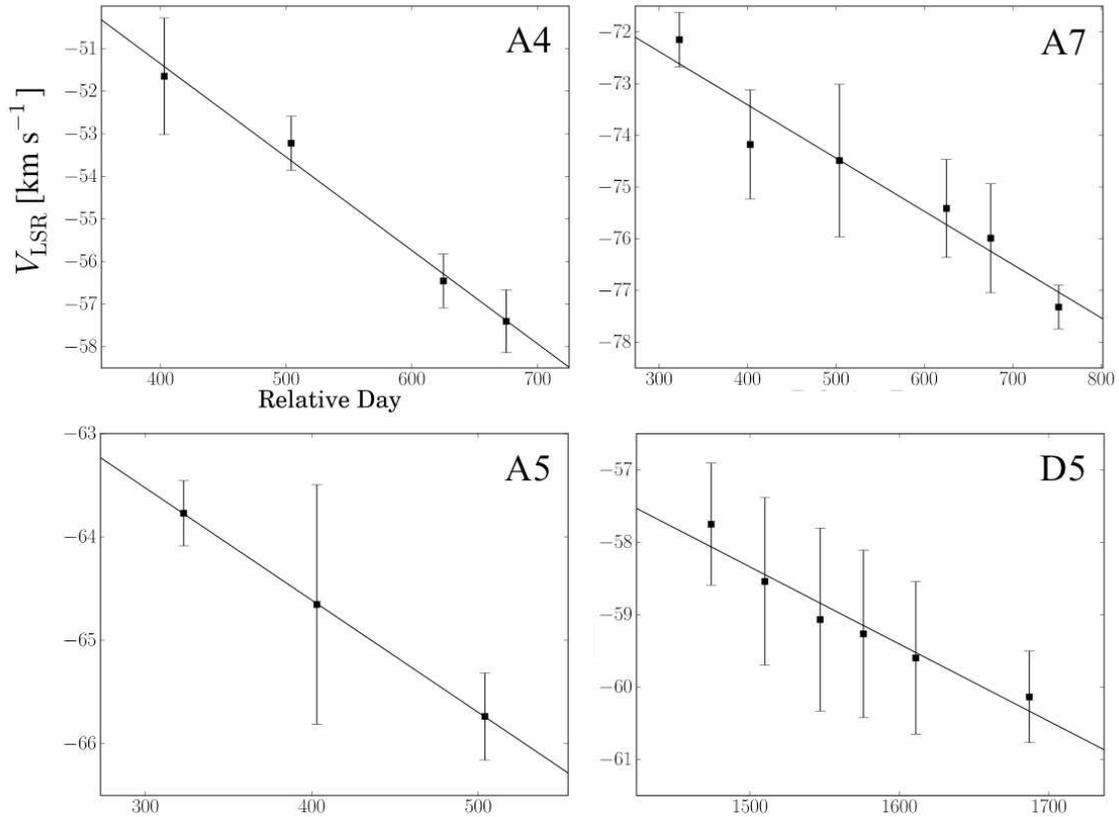}
\end{center}
 \caption{Accelerations in individual features. 
Error bars indicate a velocity dispersion in each feature. 
Black lines present a linear acceleration determined by a dispersion-weighted fitting. 
Four features (A4, A5, A7 and D5) have shown a clear signature of an acceleration beyond the velocity dispersions. 
} \label{fig:acc_feature}
\end{figure*}

 These accelerations along the LOS were also recognized in four individual maser features. 
We accepted only an acceleration that exceeds a velocity dispersion of each maser feature, in order to avoid apparent accelerations that can be caused by an internal structure of features. 
Figure \ref{fig:acc_feature} presents observed accelerations in the four features and Table \ref{tab:acceleration} summarizes linearly-fitted acceleration rates. 
The averaged acceleration rate well agrees with that derived from the dynamic spectrum. 
It should be noted that no significant acceleration is found in proper motions. 
This is probably because the acceleration of -5 km s$^{-1}$ yr$^{-1}$ is relatively small, and also, an accelerating force is mainly along the LOS. 
We will discuss about proper motions and an inclination angle of the maser outflow in the following sections. 
These accelerations along the LOS probably contribute to the short maser lifetime in G353, decreasing the coherent path for maser amplification.

\begin{figure}[!htb]
\begin{center}
\FigureFile(90mm,90mm){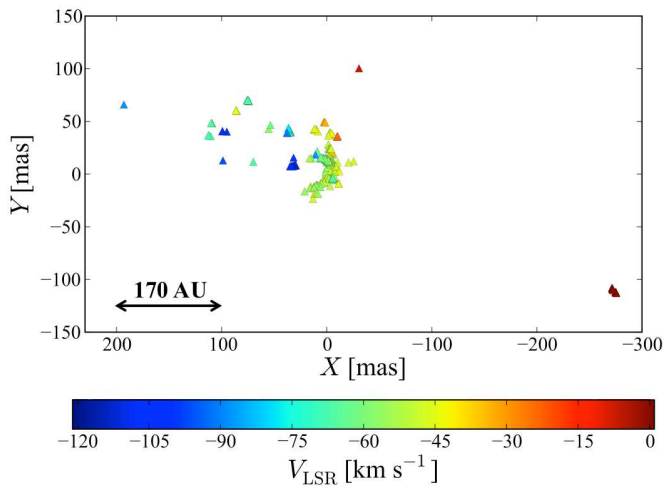}
\end{center}
 \caption{Overall distribution of all maser features detected by VERA, including that reported in Paper I. 
Filled triangle presents a position of each maser feature with color indicating a LOS velocity. 
The coordinate origin is ($\alpha$,$\delta$)$_{\rm J2000.0}$ = (\timeform{17h26m01.5883s}, \timeform{-34D15'14.903"}). } \label{fig:map}
\end{figure}

\subsection{Maser Distributions}
Figure \ref{fig:map} presents the overall distribution of maser features detected in all the VERA observations, including that reported in Paper I. 
The $X$, $Y$ and color show the right ascension offset ($X$ = $\Delta\alpha$cos$\delta$), declination offset ($Y$ = $\Delta\delta$) and LOS velocity ($V_{\rm LSR}$), respectively. 
The coordinate origin is the position of the brightest maser spot detected in Paper I, i.e., ($\alpha$,$\delta$)$_{\rm J2000.0}$ = (\timeform{17h26m01.5883s}, \timeform{-34D15'14.903"}). 

Most of the maser features are concentrated near the origin, belonging to the main velocity components (-65 -- -45 km s$^{-1}$). 
All of other significantly blue-shifted features (-120 -- -70 km s$^{-1}$) spread eastward. 
On the other hand, slightly red-shifted components ($\sim$ 0 km s$^{-1}$) were appeared at the western end during the latter 7 epochs. 
This is the first VLBI detection of red-shifted components in G353. 
We note that the most red-shifted component of +87 km s$^{-1}$ found in CP08 was out of our velocity coverage. 

\begin{figure*}[htb]
\begin{center}
 \FigureFile(140mm,140mm){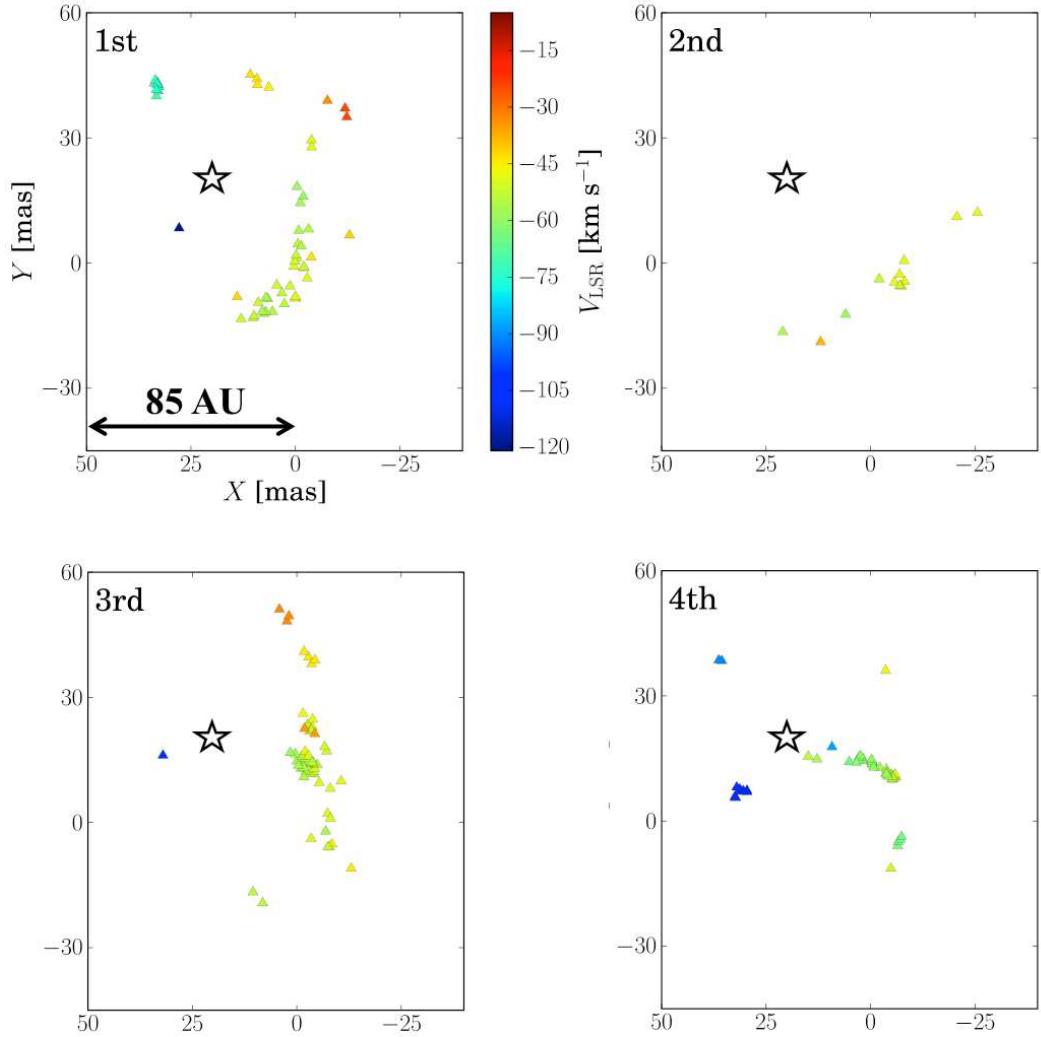}
\end{center}
 \caption{Structures of the central maser cluster in each maser flare. 
 The coordinate origin is same as that in Figure \ref{fig:map}. 
 The open star marks the peak position of the radio jet at 22 GHz, which contains a positional error of $\sim$ 50 mas, 
 including both an astrometric error reported in \citet{Motogi2013} and an internal error due to thermal image noises. 
A filamentary distribution was always found within 100 $\times$ 100 AU$^{2}$ region around the origin. 
 } \label{fig:flare}
\end{figure*}

We compared distributions of the central maser cluster appeared in each flare-up phase in Figure \ref{fig:flare}. 
Here 1st to 4th flares correspond epoch 2 -- 4, 8 -- 9, 11 -- 15 and 18 -- 23, respectively. 
We always detected filamentary maser distributions in all flares, suggesting a shock propagation. 
Although all the flares showed clearly different morphologies each other, the maser filaments always appeared in 100 $\times$ 100 AU$^{2}$ region around the origin. 

\subsection{Annual Parallax and Absolute Proper Motions}
We have successfully traced several relatively-stable maser features during 200 days in the 4th flare, thanks to the short observing interval. 
We, hence, tried to measure absolute proper motions and annual parallax of these maser features. 
We only used features that were successively detected in more than five epochs. 
Some features were also detected in epoch 17, which was observed by only three stations because of the antenna trouble in Mizusawa station. 
However, data in this epoch were not included for the astrometric analysis, because all features showed large and systematic jump ($\sim$ 3 mas) of absolute positions. 

There were five maser features available in these criteria (see Table \ref{tab:feature}). 
It should be noted that we only used $X$ data for a parallax fitting, since $Y$ data contributed not at all. 
This is because the synthesized beams were significantly elongated along the N-S direction, causing large errors in $Y$ direction (see Table \ref{tab:obs}). 
In addition, expected amplitude of a sinusoidal motion in $Y$ direction was five times smaller than that in $X$, because of a small ecliptic latitude ($\sim$ -11$^{\circ}$). 

\begin{figure}[!tb]
\begin{center}
\FigureFile(90mm,90mm){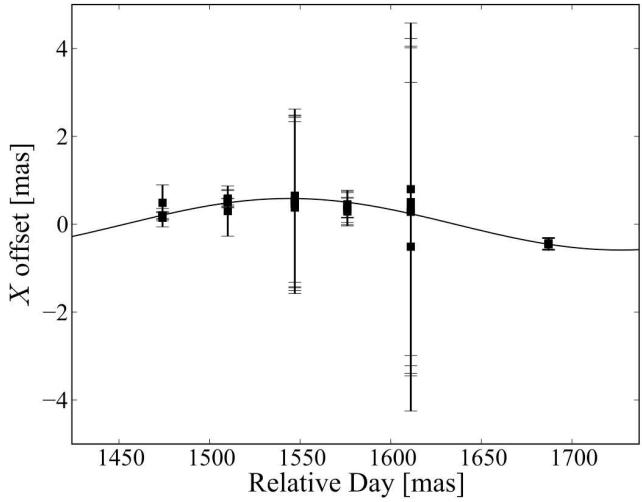}
\end{center}
 \caption{The best-fit parallax. 
All the data points used for the fitting are plotted as filled squares after subtraction of the linear absolute proper motions.  
 } \label{fig:parallax}
\end{figure}

We performed $\chi^{2}$-fitting of the common annual parallax ($\pi$) and linear absolute proper motions ($\mu^{abs}_{X}$) of each features. 
We included both of total astrometric errors in Table \ref{tab:obs} and positional dispersions in Table \ref{tab:feature} as a fitting weight. 
Figure \ref{fig:parallax} shows the best-fit parallax with all data points after the subtraction of linear proper motions. 
Individual fitting results for each maser feature are shown in supplementary Figure 2. 
The best-fit $\pi$ was 0.59 $\pm$ 0.05. 
The formal error was estimated as the reduced $\chi$ square to be unity. 

\begin{figure}[!tb]
\begin{center}
\FigureFile(90mm,90mm){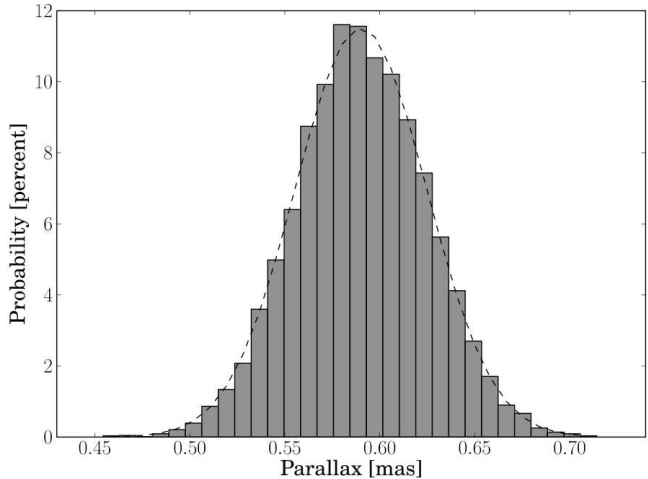}
\end{center}
 \caption{The probability distribution obtained by the simulated $\pi$ fittings in our Montecarlo analysis. 
 The dotted-curve shows the best fit gaussian distribution. 
 } \label{fig:prob}
\end{figure}

Unfortunately, a short lifetime of the maser features prevented us to trace a full parallactic cycle. 
Our parallax fitting can contains a systematic error that cannot be evaluated directly. 
We deduced such a systematic error via the Montecarlo analysis, where 10$^{4}$-times simulated fittings were performed, using the same observing epochs and the parallax of 0.59 mas. 
Positional errors in each epoch were randomly added, following a gaussian distribution with standard deviations same as the astrometric errors in Table \ref{tab:obs}. 
Figure \ref{fig:prob} presents a probability distribution of simulated $\pi$ values. 
The best-fit 1-$\sigma$ deviation of 0.035 mas was adopted as a systematic error of our fitting. 

We finally obtained a parallax of 0.59 $\pm$ 0.06, considering a root sum square of the formal fitting error and the systematic error. 
The newly derived distance is 1.70 $^{+0.19}_{-0.16}$ kpc. 
This is consistent with the commonly used photometric distance \citep{Neckel1978}. 
We, thus, conclude that our parallax measurement is well reliable and new distance of 1.7 kpc were employed in this paper. 

The best-fit absolute motions are listed in Table \ref{tab:absolute}, where absolute proper motions in $Y$ direction were fitted by a linear motions under the fixed $\pi$. 
We also determined absolute motions of three maser features, which had been excluded from the parallax fitting, in the same way. 

\subsection{Internal Proper Motions}
All the absolute proper motions determined in previous section includes a common systemic velocity that consists of the solar motion with respect to the LSR, the differential of the galactic rotation and the unknown peculiar motion of the system. 
Therefore, we also estimated relative proper motions with respect to the reference maser features, in order to discuss intrinsic 3D kinematics of the masers. 
We employed four distinct reference features, since no identical feature survived throughout all the observing epochs.  
Data were, hence, divided into four durations (epoch 2 -- 7, 8 -- 11, 12 -- 15 and 16 -- 23), depending on a reference feature. 
We estimated relative proper motions for each duration, using only a feature that detected in three or more epoch. 
All the features used for relative motion measurements are also listed in Table \ref{tab:feature}. 

Internal proper motions without any systemic motion were derived using these relative motions, as follows.  
An relative proper motion vector of i-th feature ($\vec{\mu'_{\rm i}}$) can be expressed as,  
\begin{eqnarray}
\vec{\mu'_{\rm i}} = \vec{\mu_{i}} - \vec{\mu_{0}}. 
\end{eqnarray}
Here $\vec{\mu_{i}}$ and $\vec{\mu_{0}}$ indicates an internal proper motion vector of i-th and reference maser feature, respectively.   
One can reconstruct $\vec{\mu_{i}}$, if both of $\vec{\mu^{'}_{\rm i}}$ and $\vec{\mu_{0}}$ are avilable. 
For this purpose, we deduced $\vec{\mu_{0}}$, assuming a symmetric distribution of $\vec{\mu_{i}}$ in each duration, i.e., 

\begin{eqnarray}
\sum_{i \neq 0}^{n}\vec{\mu_{i}} = 0. 
\end{eqnarray}
Here $n$ is total number of maser features except for a reference feature. In this case, equation (2) simply gives, 
\begin{eqnarray}
\vec{\mu_{0}} = -\frac{1}{n}\sum_{i \neq 0}^{n}\vec{\mu^{'}_{i}}. 
\end{eqnarray} 
We listed originally measured $\vec{\mu'_{i}}$ and reconstructed $\vec{\mu_{i}}$ in Table \ref{tab:internal}. 

\begin{figure}[!t]
\begin{center}
 \FigureFile(80mm,80mm){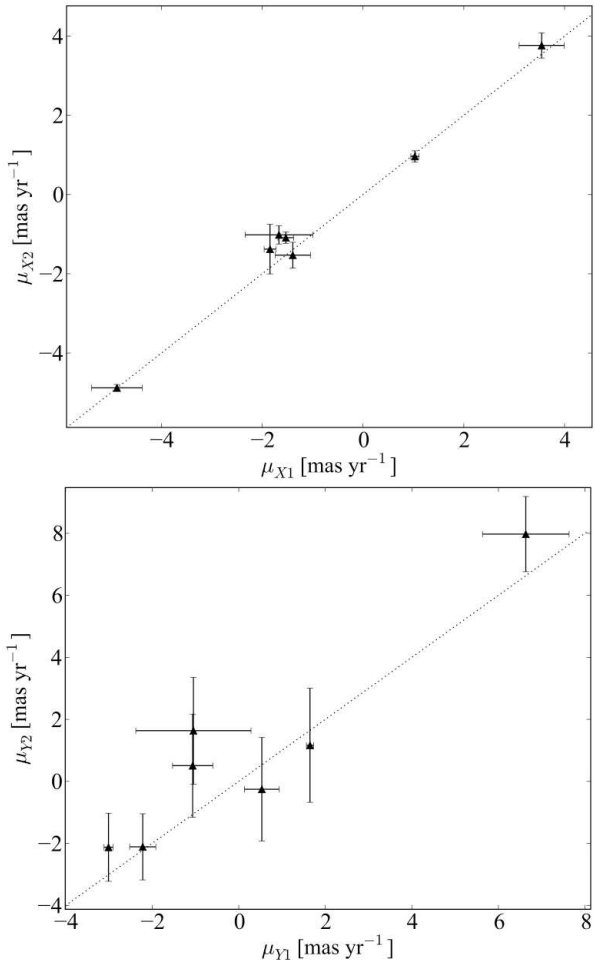}
\end{center}
 \caption{Comparison of internal proper motions derived from two different methods. 
 The horizontal axes ($\mu_{X1}$, $\mu_{Y1}$) show internal motions converted from the internal proper motions, 
 while the vertical axes ($\mu_{X2}$, $\mu_{Y2}$) show that from the absolute proper motions. 
 Two dotted lines indicate completely identical cases ($y$ = $x$). 
 } \label{fig:propers}
\end{figure}

It should be noted that the assumption of the equation (2) is adequate in our case. 
This was examined by following procedures, comparing $\vec{\mu_{i}}$ with that independently deduced from the absolute proper motions ($\vec{\mu}^{\rm abs}_i$) in Table \ref{tab:absolute}. 
We first estimated the systemic motion using the reference feature D8 in the final duration as, 
\begin{eqnarray}
\vec{\mu}_{\rm sys} = \vec{\mu}^{\rm abs}_{\rm D8} - \vec{\mu}_{\rm D8}. 
\end{eqnarray} 
Obtained $\vec{\mu}_{\rm sys}$ is 0.47 $\pm$ 0.07 and 0.99 $\pm$ 1.04 mas yr$^{-1}$ in $X$ and $Y$ direction, respectively.   
Then $\vec{\mu}^{\rm abs}_i$ were converted to $\vec{\mu_{i}}$, subtracting the $\vec{\mu}_{\rm sys}$. 
Figure \ref{fig:propers} presents a comparison of $\vec{\mu_{i}}$ determined by two different ways. 
A good correlation between two data suggests that the equation (2) is well applicable in G353. 

\begin{figure}[!tb]
\begin{center}
 \FigureFile(90mm,90mm){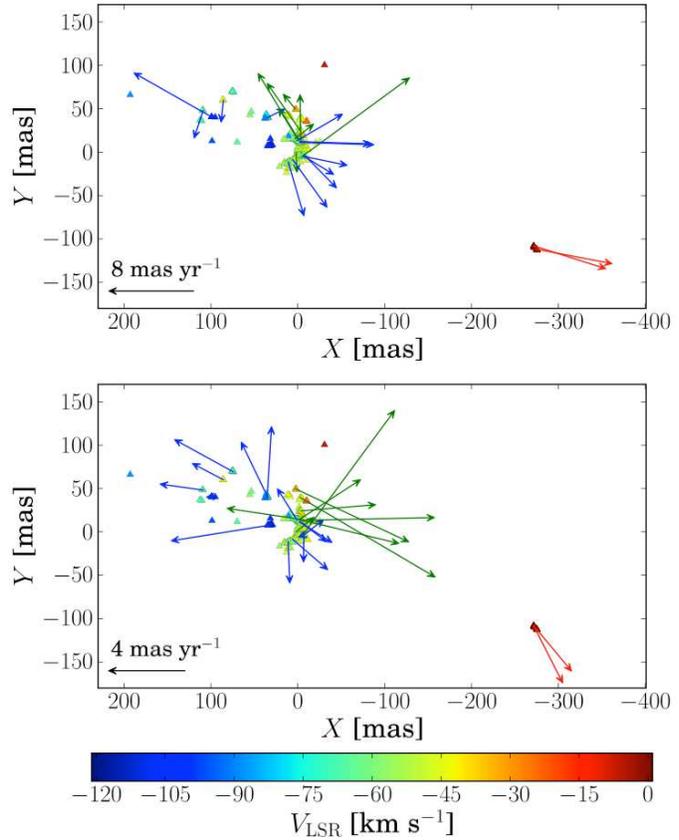}
\end{center}
 \caption{Upper panel: relative proper motions superposed on the maser distribution.  
 The blue and red arrows indicate proper motions of blue-shifted and red-shifted features, respectively. 
 The green arrows are the same as the blue ones, but with an inclination angle larger than 25$^{\circ}$ from the LOS.  
 Lower panel: internal proper motions converted from the relative proper motions, using the motion of reference maser feature. 
 } \label{fig:arrow}
\end{figure}

Figure \ref{fig:arrow} shows both relative and internal motion vectors, superposed on the maser distribution. 
The highly blue-shifted features in the east region and the red-shifted features at the west end were moving away each other, suggesting a bipolar motion along the E-W direction.  
On the other hand, the central cluster basically showed westward motions in all the durations. 
We note that this is only a projected motion and most of the 3D vectors in the cluster are nearly along the LOS (see Section 4). 

\section{Discussions}
\subsection{Inclination and Structure of the Maser Jet}
Table \ref{tab:inclination} contains 3D velocities and inclination angles, which were calculated from the internal proper motions and LOS velocities assuming the distance of 1.7 kpc. 
Almost all the blue-shifted features have 3D motions nearly along the LOS. 
This fact confirmed that the H$_{2}$O masers in G353 were associated with a face-on protostellar jet, as proposed by CP08. 
The westward expansions seen in the central cluster can be explained by a finite opening angle and face-on geometry of the jet. 

\begin{figure}[hb]
\begin{center}
 \FigureFile(85mm,85mm){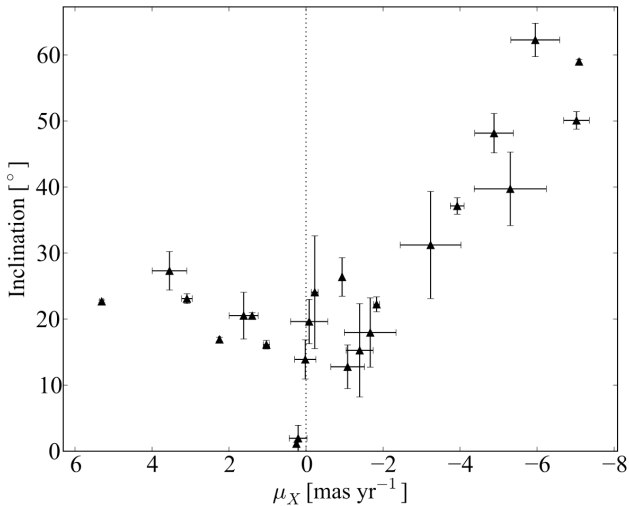}
\end{center}
 \caption{Relation between inclination angles and internal proper motions in $X$ direction.  
 The dotted line indicates $\mu_{X}$ = 0. 
 The eastward features clearly show smaller inclinations than the westward features. 
 } \label{fig:inclination}
\end{figure}

Figure \ref{fig:inclination} shows the relation of the inclination angles to the internal proper motions in $X$ direction. 
The features moving eastward are more closely along the LOS than that moving westward. 
The averaged inclination angles are 17$^{\circ}$ and 34$^{\circ}$ from the LOS for the eastward and westward features, respectively. 
If the masers are excited on the working surface between the jet and cavity wall having a constant opening angle,  
the jet axis is rather inclined ($\sim$ 8$^{\circ}$) to the west from the LOS. 

\begin{figure*}[ht]
\begin{center}
 \FigureFile(145mm,145mm){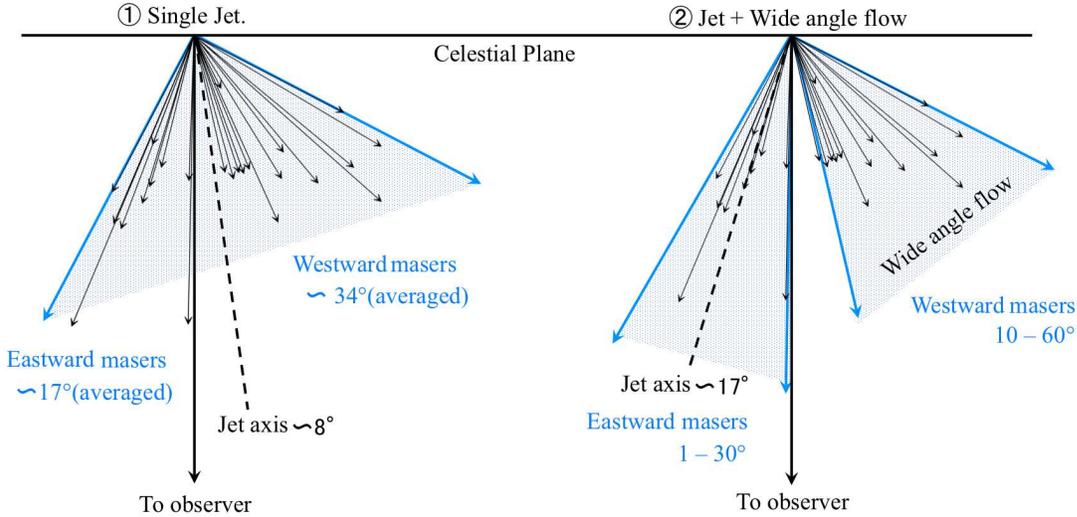}
\end{center}
 \caption{Schematic views of possible geometries in G353. 
Thick black lines indicate the LOS and celestial plane. 
Thin black arrows show inclinations and 3D velocities of all the blue-shifted masers. 
Blue hatched regions show a jet (and/or outflow) cavity. 
We also showed roughly estimated inclinations from the LOS. 
 Color version of this figure is available in on-line.  
 } \label{fig:schematic}
\end{figure*}

Another possible explanation is that the masers are associated with two distinct outflows having a different collimation as in the case of CepA HW2 \citep{Torrelles2011}, 
i.e., collimated jet and wide angle (or equatorial) outflow. 
If this is the case, the eastward features, which have rather higher 3D velocities,  may trace the collimated jet. 
A jet inclination of $\sim$ 17$^{\circ}$ is simply expected from the averaged inclination of the eastward masers. 
On the other hand, a wide angle outflow traced by the westward features would have a large opening angle up to $\sim$ 80$^{\circ}$ from the jet axis. 
Schematic views of these two cases are showed in Figure \ref{fig:schematic}. 

Although the red-shifted features at the west end strongly suggest a bipolar morphology along the E-W direction, 
this position angle seems to be perpendicular to that of the radio jet in the arcsecond scale \citep{Motogi2013}. 
This can be explained by the projection effect, or the radio jet may trace an ionized equatorial wind reported in \citet{Gibb2007} and \citet{Maud2013}. 

These red-shifted features are moving almost along the celestial plane, contrary to the blue-shifted features. 
We suggest that the blue-shifted masers trace more intrinsic outflow kinematics at the root region, 
since the peak position of the radio jet indicates that the host object is located at the almost same position as the central blue-shifted cluster (see Figure \ref{fig:flare}). 
If the red-shifted features are excited by the same driving source, the jet may be strongly bending within 500 AU from the source, 
otherwise, these features can be associated with a wide-angle outflow, discussed above.  
A VLBI position and proper motion of the extremely red-shifted component will allow us to solve these questions, 
because such high-velocity component probably reflects an intrinsic motion of the westward jet.  
An interferometric imaging of the thermal molecular jet found in \citet{Motogi2013} will be also helpful to understand the overall extent and trajectory of the jet.

\subsection{Recurrent Shock Propagations}
The recurrent formations of a filamentary maser distribution indicates episodic shock propagations, as suggested in Paper I. 
The short maser lifetime in G353 can simply emphasize one-to-one relation between maser flares and shock propagations. 
Paper I discussed that such a recurrent shock is possibly caused by an episodic jet-launching or precession. 
Although our new data do not exclude both cases, no precession signature in the radio jet may support the episodic jet-launching. 

The spatial scale in which the maser filament recurrently appeared is $\sim$ 100 $\times$ 100 AU$^{2}$. 
This scale can be resolved by the maximum resolution of Atacama Large Millimeter/Submillimeter Array (ALMA) in near future. 
Therefore, there will be a chance to directly determine which case is correct, by monitoring of how shocked gas varies using a thermal tracer. 
Such a monitoring was already succeeded in the case of the molecular jet associated with a nearby low mass protostar \citep{Jimenez2011}. 

\subsection{Jet Accelerating Region}
Figure \ref{fig:dynamic} indicates that the acceleration of the main velocity component also occurs recurrently. 
In addition, the acceleration had clearly continued beyond a lifetime of individual maser feature. 
These facts suggest that the H$_{2}$O maser in G353 traces a region where molecular gas is steadily accelerated. 
The spatial distribution of accelerating velocity components is almost consistent with the region where maser features are recurrently formed, i.e., $\sim$ 100 $\times$ 100 AU$^{2}$ wide around the root of the bipolar jet. 
If we adopt the expected inclinations and opening angles, the LOS extent of such an accelerating region is 180 -- 300 AU. 

The thermal SiO spectrum reported in \citet{Motogi2013} has indicated 
that the maximum LOS velocity of shocked molecular gas is $\sim$ -120 km s$^{-1}$, similar to that of the maser emission. 
An important fact is that such highest-velocity maser components have shown no significant acceleration in our monitoring. 
It can be interpreted as that the molecular jet is accelerated up to this velocity, and then, turns into free expansion. 
In order to examine this possibility, we must study velocity structures of both pre-shock and post-shock gas along the jet axis in multi-scale. 

The most plausible origin of such an acceleration is momentum supply from an optical jet. 
A momentum rate of the maser flow in G353 was estimated as $\sim$ 10$^{-3}$ $M_{\odot}$ km s$^{-1}$ yr$^{-1}$ in Paper I. 
This order of momentum rate is often found in high mass star formation (e.g., \cite{Wu2004}). 
If we consider a radiation pressure, a bolometric luminosity higher than 10$^{5}$ $L_{\odot}$ is required to achieve the momentum rate at the radius of 150 AU, although such a high luminosity is quite rare and may be unlikely case in G353. 
Another possible origin is a magnetic pressure gradient along the jet axis (e.g., \cite{Machida2008}). 
This can be examined by a first-principle MHD simulation, tracing rather longer-term outflow evolution than that commonly reported at present. 

\subsection{The origin of DBSMs}
\citet{Breen2010} have shown that the number of DBSMs without any OH maser (i.e., youngest evolutionary stage) is statistically superior to red-shifted counterparts where maser fluxes dominated by highly red-shifted emissions. 
Therefore, there must be some intrinsic origin of the blue-shift dominance. 
\citet{Motogi2013} have proposed that an optically thick disk in a nearly face-on geometry can selectively cause a blue-shift dominance, masking any red-shifted emission. 
This scenario is still plausible, because there is no red-shifted component around the center of the bipolar distribution, where the host HMYSO is thought to be embedded with a disk. 
Further test will be given by an exact position of the extremely red-shifted component at +87 km s$^{-1}$. 

Alternatively, if a circumstellar radiation field is completely dominated by spherically-symmetric fluxes from a central object, 
only blue-shifted components can receive seed photons for maser amplifications towards the observer. 
The two-outflow model may prefer this explanation, allowing a lack of eastward red-shifted masers associated with the wide angle outflow. 
This can be applicable in the case of G353 that has no background HII region. 
In either case, more statistical samples are required to find out the true origin. 
One must compare the geometry of a disk/jet system, optical depth of a disk and circumstellar environment in several DBSMs, 
combining a VLBI array and the highest-resolution connected array such as ALMA.  

\section{Conclusions}
We have conducted a follow-up VLBI and single-dish monitoring for the highly variable H$_{2}$O masers in G353. 
Thanks to the denser monitoring interval than that in Paper I, several properties of G353 have been revealed as follows. 

\begin{description}
\item [ (1)]  The H$_{2}$O masers are associated with the bipolar jet along the E-W direction. 
An inclination angle of the jet axis is expected to be 8$^{\circ}$ -- 17$^{\circ}$, suggesting a nearly face-on geometry as proposed in CP08.  
\\
\item [ (2)] Intermittent flare activities are caused by recurrent shock propagations. 
Such shocks are probably explained by episodic jet-launchings.  
This will be verified by a direct monitoring of shocked gas with ALMA, using a thermal tracer.  
\\
\item [ (3)] The systematic accelerations were recurrently detected and continued beyond a lifetime of individual maser features. 
These accelerating masers probably trace a region, where molecular gas is steadily accelerated up to -120 km s$^{-1}$ within 100$\times$100$\times$300 AU$^{3}$ box ($X$, $Y$ and the LOS, respectively) around the root of the bipolar jet. 
\\
\item [ (4)] The measured annual parallax indicates a distance of 1.70 $^{+0.19}_{-0.16}$ kpc, supporting a commonly-used photometric distance. 
\\
\item [ (5)] The origin of DBSMs is still open question. This can be explained by a masking effect of optically thick face-on disk as proposed in Paper I.  
We require more statistical data about a geometry of a disk/jet system, optical depth of a disk and circumstellar environment towards the known DBSMs. 

\end{description}


\bigskip

The author heavily thanks to all the members in VERA and Hokkaido Univerisity Tomakomai Radio Observatory 
for their assistance in operations on the very long-term monitoring. 
The author also thanks to the anonymous referee for useful suggestions and very encouraging comments. 
This work was financially supported by the Grant-in-Aid for the Japan Society for the Promotion of Science Fellows (K.M.) 
and the Grants-in-Aid by the Ministry of Education, Culture, and Science of Japan 24-6525 (K.M.). 

\appendix
\section{Tables}

\begin{table*}[!hb]
  \caption{Summary of VERA Obsevations}\label{tab:obs}
   \begin{tabular}{cccccccl}\hline
   Epoch& Day$^{a}$& $\theta_{\rm b}$ & PA & $\Delta I$$^{b}$ &$\sigma_{1}$ & $\sigma_{2}^{c}$ & Comments$^{d}$ \rule[0mm]{0mm}{4mm}\\
       && (mas $\times$ mas) & ($^{\circ}$) & (Jy beam$^{-1}$) & (mas $\times$ mas) & (mas $\times$ mas)  & \\ \hline
1 & 183 & 3.16 $\times$ 0.74 & -17.1  & 0.48  & 0.15 $\times$ 0.49 & - & - \\
2 & 323 & 3.98 $\times$ 0.82 & -23.2  & 0.30  & 0.92 $\times$ 2.15 & - & $T_{\rm sys}$ $\sim$ 1000 K at OG \\
3 & 403 & 2.63 $\times$ 0.91 & -14.1  & 0.37  & 0.06 $\times$ 0.24 & - & - \\
4 & 504 & 2.57 $\times$ 0.88 & -13.2  & 0.30  & 0.04 $\times$ 0.16 & - & - \\
5 & 625 & 2.86 $\times$ 0.82 & -12.3  & 0.37  & 0.11 $\times$ 0.48 & - & - \\
6 & 675 & 2.79 $\times$ 0.79 & -13.3  & 0.16  & 0.23 $\times$ 0.98 & - & - \\
7 & 752 & 3.15 $\times$ 0.85 & -19.5  & 0.21  & 0.09 $\times$ 0.25 & - & - \\
8 & 842 & 2.59 $\times$ 0.75 & -14.1  & 0.40  & 0.10 $\times$ 0.41 & - & - \\
9 & 976 & 2.84 $\times$ 0.76 & -13.8  & 0.29  & 0.20 $\times$ 0.80 & - & $T_{\rm sys}$ $\sim$ 2000 K at OG \\
10 & 1036 & 2.79 $\times$ 0.77 & -15.0  & 0.19  & 0.21 $\times$ 0.77 & - & $T_{\rm sys}$ $\sim$ 1500 K at IS \\
11 & 1069 & 2.92 $\times$ 0.82 & -14.5  & 0.18  & 0.36 $\times$ 1.40 & 0.03 $\times$ 0.02 & - \\
12 & 1107 & 2.67 $\times$ 0.91 & -7.3  & 0.23  & 0.03 $\times$ 0.22 & 0.28 $\times$ 0.00 & - \\
13 & 1129 & 2.56 $\times$ 0.83 & -11.7  & 0.14  & 0.16 $\times$ 0.75 & 0.00 $\times$ 0.02 & - \\
14 & 1164 & 2.65 $\times$ 0.85 & -10.5  & 0.11  & 0.09 $\times$ 0.50 & 2.44 $\times$ 1.81 & - \\
15 & 1210 & 3.03 $\times$ 0.78 & -16.9  & 0.20  & 0.52 $\times$ 1.72 & 0.08 $\times$ 0.11 & - \\
16 & 1320 & 2.84 $\times$ 0.78 & -10.0  & 0.48  & 0.15 $\times$ 0.88 & 0.10 $\times$ 0.02 & - \\
17 & 1398 & 2.66 $\times$ 0.66 & -15.2  & 0.26  & 0.33 $\times$ 1.19 & 0.21 $\times$ 0.00 & Three stations without IR \\
18 & 1474 & 2.76 $\times$ 0.85 & -11.1  & 0.16  & 0.07 $\times$ 0.33 & 0.00 $\times$ 0.00 & - \\
19 & 1510 & 2.60 $\times$ 0.82 & -12.8  & 0.17  & 0.03 $\times$ 0.13 & 0.00 $\times$ 0.00 & - \\
20 & 1547 & 2.54 $\times$ 0.80 & -11.8  & 0.23  & 0.11 $\times$ 0.48 & 1.95 $\times$ 2.65 & - \\
21 & 1576 & 2.61 $\times$ 0.84 & -10.0  & 0.22  & 0.10 $\times$ 0.52 & 0.26 $\times$ 1.75 & - \\
22 & 1611 & 3.27 $\times$ 0.83 & -15.7  & 0.37  & 0.22 $\times$ 0.78 & 3.72 $\times$ 0.06 & $T_{\rm sys}$ $\sim$ 1000 K at OG \\
23 & 1687 & 2.59 $\times$ 0.76 & -4.7  & 0.58  & 0.10 $\times$ 1.31 & 0.07 $\times$ 0.06 & $T_{\rm sys}$ $\sim$ 2000 K at MZ and IR \\ \hline
\multicolumn {8} {l} {$^a$ The relative days counted from January 1st, 2008, that can be converted to MJD by adding 54465. }\\
\multicolumn {8} {l} {$^b$ Typical value in self-calibrated images.}\\
\multicolumn {8} {l} {$^c$ Only available in the epoch 11 to 23, where we employed $\sigma$ = $\sqrt{\sigma_{1}^2+\sigma_{2}^2}$ as a total error.} \\
\multicolumn {8} {l} {$^d$ MZ, IR, IS, OG: Mizusawa, Iriki, Ishigaki, Ogasawara station, respectively.} \\
     \end{tabular}
\end{table*}

\begin{longtable}[!hb]{cc|cc|cc|cc|c}\hline
\caption{The features used for kinematic analyses}\label{tab:feature}\hline
ID & Epoch &$V_{LSR}$ & $\Delta$$V^{a}$  &$X^{b, c, d}$ & $Y$ & $X$ - $X_{0}$$^{d}$ &  $Y$ - $Y_{0}$ & $S_{\rm int}$ \\
& & \multicolumn{2}{|c|}{(km s$^{-1}$)} & \multicolumn{2}{|c|}{(mas)}& \multicolumn{2}{c|}{(mas)}  & (Jy km s$^{-1}$) \\ \hline
\endfirsthead
\hline
ID & Epoch &$V_{LSR}$ & $\Delta$$V^{a}$ &$X^{b, c}$ & $Y$ & $X$ - $X_{0}$$^{d}$ &  $Y$ - $Y_{0}$ & $S_{\rm int}$ \\
& &  \multicolumn{2}{|c|}{(km s$^{-1}$)} & \multicolumn{2}{|c|}{(mas)}& \multicolumn{2}{c|}{(mas)}  & (Jy km s$^{-1}$) \\ \hline
\endhead
\hline
\endfoot
\hline 
\multicolumn {9} {l} {$^a$ Total velocity widths (full width at zero intensity). }\\
\multicolumn {9} {l} {$^b$ All the parenthetic values indicate a error. }\\
\multicolumn{9}{l}{$^c$ All features also contain the relevant astrometric error in Table \ref{tab:obs}}\\
\multicolumn {9} {l} {$^d$ The coordinate origin is ($\alpha$,$\delta$)$_{\rm J2000.0}$ = (\timeform{17h26m01.5883s}, \timeform{-34D15'14.903"}).}\\
\multicolumn{9}{l}{$^e$ Column 7, 8 show positional offsets from the referenced features. }\\
\endlastfoot
\hline \hline
\multicolumn{9}{c}{epoch 2 -- 7 (referenced feature: A7))}\\ \hline \hline
A1 & 3 & -23.72 & 1.26 & -11.90 (0.14) & 37.09 (0.04)& -44.96 & -5.80 & 2.63 (0.11) \\
 &  4 & -23.78 & 1.68 & -12.31 (0.18) &35.01 (0.13)& -45.62 & -6.54 & 7.79 (0.21) \\
 &  6 & -24.13 & 0.21 & -12.84 (0.04) &34.28 (0.14)& -46.70 & -7.34 & 0.25 (0.05) \\ \hline
A2 & 4 & -43.83 & 0.42 & 83.35 (0.07) &59.54 (0.00)& 50.03 & 17.99 & 0.93 (0.10) \\
 &  5 & -43.79 & 0.63 & 83.17 (0.08) &58.34 (0.09) & 50.19 & 17.27 & 1.51 (0.19) \\
 &  6 & -43.82 & 0.84 & 83.83 (0.15) &58.68 (0.14) & 49.97 & 17.06 & 1.64 (0.10) \\ \hline
A3 & 2 & -51.44 & 5.69 & 4.46 (0.31) &-5.32 (0.33) & -29.05 & -49.20 & 1023.54 (9.40) \\
 & 3 & -53.08 & 3.79 & 3.24 (0.41) &-7.18 (0.28)  & -29.82 & -50.06 & 824.36 (2.38) \\
 &  4 & -53.54 & 1.90 & 2.65 (0.09) &-9.83 (0.06)  & -30.67 & -51.38 & 16.96 (0.63) \\ \hline
A4 & 3 & -51.65 & 2.74 & 8.90 (0.39) &-9.47 (0.37) & -24.16 & -52.35 & 35.59 (1.48) \\
 & 4 & -53.22 & 1.26 & 9.91 (0.10) &-12.67 (0.03)  & -23.41 & -54.22 & 19.55 (0.46) \\
  & 5 & -56.46 & 1.26 & 9.10 (0.23) &-14.84 (0.16) & -23.88 & -55.90 & 3.53 (0.31) \\
  & 6 & -57.41 & 1.47 & 9.64 (0.06) &-14.74 (0.33) & -24.22 & -56.36 & 3.59 (0.14) \\ \hline
A5 & 2 & -63.77 & 0.63 & 107.57 (0.00) &51.48 (0.04) & 74.06 & 7.60 & 1.87 (0.18) \\
 & 3 & -64.65 & 2.32 & 107.25 (0.14) &50.07 (0.24) & 74.19 & 7.19 & 6.47 (0.17) \\
  & 4 & -65.74 & 0.84 & 107.80 (0.03) &47.82 (0.20) & 74.48 & 6.27 & 1.59 (0.13) \\ \hline
A6 & 2 & -70.00 & 1.69 & 33.79 (0.05) &43.07 (0.24) & 0.28 & -0.81 & 17.90 (0.34) \\
  & 3 & -72.43 & 2.32 & 32.89 (0.37) &42.82 (0.16)  & -0.17 & -0.07 & 28.06 (0.15) \\
  & 4 & -71.84 & 3.16 & 32.80 (0.17) &41.24 (0.23)  & -0.52 & -0.31 & 26.93 (0.31) \\ \hline
A7  & 2 & -72.15 & 1.05 & 33.51 (0.03) &43.88 (0.11)&  0.00 & 0.00 & 4.44 (0.21) \\
 & 3 & -74.18 & 2.11 & 33.06 (0.12) &42.88 (0.08)  & 0.00 & 0.00 & 13.53 (0.13) \\
  & 4 & -74.49 & 2.95 & 33.32 (0.11) &41.55 (0.08) & 0.00 & 0.00 & 146.72 (0.28) \\
  & 5 & -75.41 & 1.90 & 32.98 (0.04) &41.06 (0.32) & 0.00 & 0.00 & 19.77 (0.68) \\
  & 6 & -75.99 & 2.11 & 33.86 (0.04) &41.62 (0.04) & 0.00 & 0.00 & 24.79 (0.23 \\
  & 7 & -77.32 & 0.84 & 33.30 (0.05) &42.94 (0.12) & 0.00 & 0.00 & 1.08 (0.11) \\ \hline\hline
  \multicolumn{9}{c}{epoch 8 -- 11 (referenced feature: B4)}\\ \hline\hline
B1  & 8 & -46.75 & 1.47 & -6.99 (0.04) & -2.76 (0.37) & -80.95 & -71.78 & 10.21 (0.44) \\ 
 & 9 & -47.08 & 2.74 & -8.20 (0.05) & -4.45 (0.07) & -82.21 & -72.62 & 353.44 (0.47) \\
  & 10 & -46.86 & 2.74 & -9.24 (0.07) & -3.74 (0.06) & -82.50 & -73.20 & 55.49 (0.27) \\
  & 11 & -47.11 & 2.74 & -8.42 (0.17) & -5.18 (0.17) & -82.70 & -73.39 & 7.01 (0.21) \\ \hline
B2 & 8 & -48.48 & 2.32 & -5.72 (0.07) & -4.61 (0.50) & -79.68 & -73.64 & 62.72 (0.78) \\
  & 9 & -48.54 & 1.69 & -7.39 (0.05) & -5.58 (0.22) & -81.39 & -73.76 & 55.04 (0.34) \\
  & 10 & -48.79 & 1.69 & -8.63 (0.13) & -4.50 (0.25) & -81.89 & -73.96 & 13.10 (0.21) \\
  & 11 & -49.24 & 1.90 & -7.83 (0.11) & -5.84 (0.10) & -82.11 & -74.05 & 5.19 (0.16) \\ \hline
B3 & 9 & -51.50 & 3.37 & -7.00 (0.21) & -5.46 (0.24) & -81.01 & -73.64 & 24.30 (0.40) \\
  & 10 & -51.88 & 2.11 & -8.17 (0.06) & -4.50 (0.28) & -81.43 & -73.95 & 12.59 (0.28) \\
  & 11 & -51.71 & 1.69 & -7.43 (0.05) & -5.89 (0.13) & -81.70 & -74.09 & 6.33 (0.28) \\ \hline
B4 & 8 & -67.03 & 0.63 & 73.96 (0.05) & 69.02 (0.04) & 0.00 & 0.00 & 1.76 (0.17) \\
 & 9 & -66.84 & 1.26 & 74.01 (0.11) & 68.18 (0.09) & 0.00 & 0.00 & 2.74 (0.19) \\
 &  10 & -66.83 & 0.84 & 73.26 (0.01) & 69.46 (0.08) & 0.00 & 0.00 & 1.45 (0.10) \\
 &  11 & -66.86 & 0.84 & 74.28 (0.09) & 68.21 (0.17) & 0.00 & 0.00 & 1.54 (0.11) \\ \hline\hline
\multicolumn{9}{c}{epoch 12 -- 15 (referenced feature: C5)}\\ \hline\hline
C1 & 13 & -32.50 & 0.63 & 2.36 (0.04) & 48.24 (0.05) & 5.91) & 10.22 & 0.44 (0.05)\\
 &  14 & -32.33 & 0.63 & 4.16 (0.20) & 51.10 (0.13) & 5.88) & 10.13 & 0.54 (0.05) \\
 &  15 & -32.51 & 1.05 & 1.88 (0.20) & 49.50 (0.12) & 6.22) & 10.57 & 0.85 (0.09) \\ \hline
C2  & 12 & -44.45 & 0.42 & -2.80 (0.01) & 39.62 (0.02) & 0.00 & 0.00 & 0.55 (0.08) \\
 &  13 & -44.49 & 1.05 & -3.55 (0.21) & 38.03 (0.24) & 0.00) & 0.00 & 1.12 (0.08) \\
 &  14 & -44.56 & 1.05 & -1.72 (0.28) & 40.97 (0.21) & 0.00) & 0.00 & 1.02 (0.07) \\
 &  15 & -44.53 & 0.63 & -4.34 (0.14) & 38.93 (0.06) & 0.00) & 0.00 & 0.72 (0.07) \\ \hline
C3 & 13 & -45.85 & 1.68 & -3.72 (0.06) & 22.83 (0.10) & -0.17 & -15.20 & 1.73 (0.09) \\
 &  14 & -46.65 & 2.32 & -1.52 (0.11) & 26.14 (0.08) & 0.20) & -14.83 & 4.10 (0.10) \\
 &  15 & -47.22 & 1.47 & -3.80 (0.04) & 24.72 (0.08) & 0.54) & -14.21 & 2.33 (0.11) \\ \hline
C4 & 12 & -51.63 & 2.32 & -3.44 (0.26) & 13.21 (0.15) & -0.64 & -26.41 & 61.82 (0.47) \\
 & 13 & -51.93 & 3.37 & -4.04 (0.32) & 12.13 (0.15) & -0.49) & -25.90 & 144.56 (0.33) \\
  & 14 & -52.26 & 2.95 & -2.14 (0.48) & 15.34 (0.24) & -0.42 & -25.63 & 123.73 (0.29) \\
  & 15 & -52.25 & 3.16 & -4.91 (0.15) & 13.80 (0.21) & -0.57 & -25.14 & 520.38 (0.67) \\ \hline
C5 & 12 & -54.86 & 1.47 & -1.30 (0.12) & 14.05 (0.10) & 1.50 & -25.57 & 25.71 (0.30) \\
 & 13 & -56.18 & 1.90 & -1.68 (0.21) & 13.18 (0.12) & 1.87) & -24.85 & 16.14 (0.16) \\
  & 14 & -56.82 & 1.68 & 0.38 (0.16) & 16.43 (0.09) & 2.10) & -24.54 & 4.63 (0.10) \\ \hline
C6 & 12 & -56.95 & 2.95 & -0.01 (0.52) & 14.73 (0.22) & 2.79 & -24.89 & 38.99 (0.32) \\
  & 13 & -57.54 & 2.11 & -0.38 (0.17) & 13.68 (0.11) & 3.17) & -24.35 & 17.58 (0.17) \\
  & 14 & -58.33 & 1.90 & 1.58 (0.11) & 16.72 (0.14) & 3.30) & -24.25 & 3.07 (0.10) \\ \hline
C7 & 12 & -107.60 & 1.05 & 99.21 (0.04) & 40.58 (0.09) & 102.01 & 0.96 & 2.21 (0.10) \\
  & 14 & -107.76 & 1.47 & 101.43 (0.02) & 42.60 (0.04) & 103.16 & 1.63 & 8.22 (0.08) \\
  & 15 & -107.68 & 2.95 & 99.77 (0.19) & 41.01 (0.19) & 104.10 & 2.08 & 16.99 (0.15) \\ \hline \hline
\multicolumn{9}{c}{epoch 17 -- 23 (referenced feature: D8)}\\ \hline\hline
D1 & 17 & -0.14 & 0.42 & -272.66 (0.02) & -113.40 (0.04) & -301.79 & -115.95 & 0.68 (0.13) \\
  & 18 & -0.08 & 0.63 & -273.74 (0.05) & -109.01 (0.16) & -303.30 & -116.29 & 1.11 (0.08) \\
  & 19 & -0.11 & 0.84 & -273.51 (0.09) & -109.26 (0.08) & -303.95 & -116.49 & 1.35 (0.14) \\
  & 20 & -0.06 & 0.42 & -273.56 (0.02) & -109.74 (0.03) & -304.61 & -116.83 & 0.85 (0.09) \\
  & 21 & -0.01 & 0.42 & -273.67 (0.10) & -109.43 (0.29) & -304.97 & -116.97 & 0.39 (0.07) \\
  & 22 & -0.05 & 0.21 & -273.85 (0.08) & -109.61 (0.25) & -305.89 & -117.69 & 0.32 (0.11) \\ \hline
D2 & 17 & -0.53 & 0.42 & -269.42 (0.05) & -109.57 (0.02) & -305.03 & -119.78 & 0.84 (0.15) \\
  & 18 & -0.36 & 0.84 & -277.09 (0.07) & -112.55 (0.06) & -306.65 & -119.82 & 2.77 (0.09) \\
  & 19 & -0.32 & 1.05 & -276.76 (0.15) & -112.71 (0.11) & -307.21 & -119.94 & 2.69 (0.14) \\
  & 20 & -0.32 & 0.84 & -276.93 (0.03) & -113.14 (0.03) & -307.98 & -120.24 & 2.21 (0.11) \\
  & 21 & -0.31 & 0.63 & -277.15 (0.05) & -112.99 (0.04) & -308.45 & -120.54 & 1.30 (0.10) \\ \hline
D3 & 17 & -53.78 & 1.90 & -0.54 (0.31) & 9.61 (0.23) & -32.91 & 3.23 & 11.60 (0.37) \\
  & 18 & -53.89 & 4.42 & -5.33 (0.66) & 10.05 (0.70) & -34.89 & 2.77 & 75.71 (0.34) \\
  & 19 & -53.96 & 3.16 & -5.22 (0.19) & 9.98 (0.19) & -35.67 & 2.75 & 40.02 (0.28) \\
  & 20 & -54.16 & 2.95 & -5.11 (0.19) & 10.07 (0.22) & -36.16 & 2.98 & 25.59 (0.34) \\
  & 21 & -53.49 & 1.90 & -5.28 (0.08) & 10.56 (0.10) & -36.59 & 3.02 & 6.14 (0.14) \\ \hline
D4 & 18 & -56.53 & 1.6 & -4.01 (0.04) & 11.49 (0.33) & -33.57 & 4.22 & 3.53 (0.15) \\
  & 19 & -56.01 & 2.74 & -3.74 (0.19) & 11.33 (0.12) & -34.19 & 4.10 & 27.48 (0.27) \\
  & 20 & -56.23 & 2.32 & -4.05 (0.10) & 10.94 (0.06) & -35.09 & 3.85 & 42.13 (0.66) \\
  & 21 & -55.98 & 2.11 & -4.08 (0.08) & 11.73 (0.14) & -35.38 & 4.18 & 23.03 (0.33) \\
  & 22 & -55.88 & 4.21 & -3.81 (0.67) & 12.41 (0.49) & -35.85 & 4.33 & 147.80 (0.53) \\ \hline
D5 & 17 & -57.10 & 0.63 & 2.73 (0.10) & 11.70 (0.07) & -29.64 & 5.32 & 2.21 (0.19) \\
  & 18 & -57.75 & 1.69 & -0.96 (0.20) & 13.13 (0.37) & -30.52 & 5.85 & 8.73 (0.18) \\
  & 19 & -58.54 & 2.32 & -0.38 (0.21) & 13.67 (0.53) & -30.83 & 6.44 & 23.56 (0.40) \\
  & 20 & -59.07 & 2.53 & -0.36 (0.26) & 13.66 (0.62) & -31.41 & 6.57 & 41.85 (0.45) \\
  & 21 & -59.27 & 2.32 & -0.34 (0.30) & 14.22 (0.29) & -31.65 & 6.68 & 58.96 (0.78) \\
  & 22 & -59.60 & 2.11 & -0.29 (0.25) & 14.71 (0.18) & -32.32 & 6.63 & 35.44 (0.79) \\
  & 23 & -60.14 & 1.26 & -0.73 (0.07) & 12.90 (0.07) & -33.01 & 7.30 & 42.06 (0.66) \\ \hline
D6 & 17 & -59.11 & 2.74 & 5.41 (0.35) & 13.70 (0.12) & -26.96 & 7.32 & 33.64 (0.38) \\
  & 18 & -60.58 & 3.16 & 1.80 (0.40) & 14.90 (0.20) & -27.76 & 7.62 & 133.79 (0.29) \\
  & 19 & -60.30 & 3.58 & 2.03 (0.57) & 14.52 (0.35) & -28.41 & 7.29 & 39.25 (0.42) \\
  & 20 & -61.21 & 2.32 & 2.81 (0.27) & 14.48 (0.13) & -28.40 & 7.29 & 17.95 (0.34) \\
  & 21 & -61.16 & 2.74 & 2.79 (0.12) & 15.23 (0.14) & -28.52 & 7.69 & 84.02 (0.29) \\
  & 22 & -60.71 & 2.53 & 2.39 (0.36) & 15.62 (0.12) & -29.64 & 7.54 & 53.42 (0.84) \\
  & 23 & -61.38 & 1.05 & 3.35 (0.04) & 14.04 (0.08) & -28.93 & 8.44 & 14.58 (0.43) \\ \hline
D7 & 18 & -63.28 & 1.47 & -6.51 (0.05) & -5.91 (0.11) & -36.07 & -13.19 & 2.22 (0.12) \\
  & 19 & -63.78 & 2.11 & -6.67 (0.23) & -5.00 (0.60) & -37.12 & -12.22 & 5.67 (0.18) \\
  & 20 & -64.63 & 2.11 & -7.12 (0.03) & -4.41 (0.09) & -38.17 & -11.50 & 13.95 (0.23) \\
  & 21 & -64.79 & 2.11 & -7.45 (0.08) & -3.76 (0.15) & -38.76 & -11.31 & 11.78 (0.15) \\ \hline
D8 & 17 & -109.78 & 0.63 & 32.37 (0.74) & 6.38 (0.02) & 0.00 & 0.00 & 1.09 (0.16) \\
 &  18 & -108.43 & 1.69 & 29.56 (0.02) & 7.28 (0.07) & 0.00 & 0.00 & 9.03 (0.11) \\
 &  19 & -108.76 & 2.95 & 30.45 (0.06) & 7.23 (0.18) & 0.00 & 0.00 & 16.16 (0.16) \\
 &  20 & -108.41 & 2.95 & 31.05 (0.08) & 7.09 (0.10) & 0.00 & 0.00 & 24.30 (0.32) \\
 &  21 & -107.93 & 2.95 & 31.30 (0.07) & 7.62 (0.34) & 0.00 & 0.00 & 21.77 (0.19) \\
 &  22 & -107.95 & 2.53 & 32.04 (0.06) & 8.08 (0.07) & 0.00 & 0.00 & 42.96 (0.38) \\
 &  23 & -107.93 & 1.69 & 32.28 (0.02) & 5.60 (0.04) & 0.00 & 0.00 & 48.64 (0.65) \\
\end{longtable}

\clearpage

\begin{table}[!htb]
\centering
  \caption{Acceleration rate along the LOS}\label{tab:acceleration}
   \begin{tabular}{lcc}\hline
$V_{\rm LSR}$/Feature & Acceleration Rate &Note \\
& (km s$^{-1}$ yr$^{-1}$) & \\ \hline
 & & \\ \hline\hline
\multicolumn{3}{c}{from the dynamic spectrum} \\ \hline\hline
 $\sim$ -50 km s$^{-1}$& -5.44  $\pm$ 0.18  & This work \\
 $\sim$ -50 km s$^{-1}$ & -5.40  $\pm$  1.20  & Paper I \\
 $\sim$ -75 km s$^{-1}$ & -4.80  $\pm$  0.50  & Paper I \\ \hline\hline
Average & -5.21 $\pm$ 0.44 & \\ \hline
& &  \\ \hline\hline
 \multicolumn{3}{c}{from the individual features} \\ \hline\hline
A5 & -3.97 $\pm$ 0.03  & This work \\
A7 & -3.76 $\pm$ 0.50  & This work \\
A4 & -8.01 $\pm$ 0.61  & This work \\
D5 & -3.89 $\pm$ 0.50  & This work \\ \hline
Average & -4.91 $\pm$ 0.15& \\ \hline
     \end{tabular} 
\end{table}

\begin{table}[!htb]
\centering
  \caption{Absolute proper motions}\label{tab:absolute}
   \begin{tabular}{ccc}\hline
   ID$^{a}$  & $\mu^{\rm abs}_{X}$ &$\mu^{\rm abs}_{Y}$\\ 
 &  (mas yr$^{-1})$  & (mas yr$^{-1}$)  \\ \hline 
D1 & -0.6 (0.1) & -1.1 (0.4)   \\
D2 & -0.9 (0.6)  & -1.1 (0.2) \\
D3 & -0.6 (0.2)  & 1.5 (1.3)    \\
D4 & -1.1 (0.3)  & 2.6 (1.4)   \\
D5 & 1.4 (0.1)  & 2.2 (1.5)   \\
D6 & 4.2 (0.3)  & 0.7 (1.3)   \\
D7 & -4.4 (0.0)  & 9.0 (0.6)   \\
D8 & 5.8 (0.1)  & 0.14 (1.47)  \\ \hline
\multicolumn{3}{l}{D2, D3 and D7 were not used for the $\pi$ fitting. }\\
\multicolumn{3}{l}{All the parenthetic values indicate an error.}\\
     \end{tabular} 
\end{table}

\begin{table}[!htb]
\centering
  \caption{Relative and internal proper motions}\label{tab:internal}
   \begin{tabular}{c|cc|cc}\hline
 & \multicolumn{2}{c|}{Relative Motions} &  \multicolumn{2}{c}{Internal Motions} \\
ID     & $\mu'_{X}$ & $\mu'_{Y}$  & $\mu_{X}$ & $\mu_{Y}$ \\
     & \multicolumn{2}{c|}{(mas yr$^{-1}$)} & \multicolumn{2}{c}{(mas yr$^{-1}$)} \\ \hline \hline
     \multicolumn{5}{c}{epoch 2 -- 7 (referenced feature: A7)}\\ \hline \hline
A1 & -2.3  (0.0) & -2.2  (0.2) & -0.9  (0.0) & 0.7  (0.3) \\
A2 & 0.2  (0.4) & -2.1  (0.1) & 1.6  (0.4) & 0.8  (0.2) \\
A3 & -3.2  (0.1) & -4.5  (0.2)  & -1.8  (0.1) & -1.6  (0.2) \\
A4 & -1.5  (0.5) & -5.1  (0.4) & -0.1  (0.5) & -2.2  (0.4) \\
A5 & 0.8  (0.0) & -2.6  (0.3) & 2.3  (0.0) & 0.3  (0.3) \\
A6 & -1.6  (0.1) & 0.8  (1.5)  & -0.2  (0.1) & 3.7  (1.5) \\
A7 & - & - & 1.4  (0.0) & 2.9  (0.1) \\ \hline \hline
   \multicolumn{5}{c}{epoch 8 -- 11 (referenced feature: B4)}\\ \hline\hline
B1 & -3.1  (0.2) & -3.1  (0.3) & 0.0  (0.3) & -1.4  (0.3) \\
B2 & -4.2  (0.4) & -0.9  (0.2)  & -1.1  (0.4) & 0.8  (0.2) \\
B3 & -2.9  (0.2) & -1.8  (0.1) & 0.2  (0.2) & -0.1  (0.1) \\
B4 & - & - & 3.1  (0.1) & 1.7  (0.1) \\ \hline\hline
   \multicolumn{5}{c}{epoch 12 -- 15 (referenced feature: C2)}\\ \hline\hline
C1 & 1.1  (0.6) & 1.4  (0.9)  & -6.0  (0.6) & -2.8  (0.9) \\
C3 & 3.2  (0.2) & 4.5  (0.3)  & -3.9  (0.2) & 0.3  (0.4) \\
C4 & 0.1  (0.3) & 4.3  (0.7) & -7.0  (0.3) & 0.2  (0.8) \\
C5 & 3.9  (0.8) & 6.3  (1.9) & -3.2  (0.8) & 2.1  (1.9) \\
C6 & 1.8  (0.9) & 2.9  (2.4)  & -5.3  (0.9) & -1.3  (2.4) \\
C7 & 7.3  (0.0) & 4.2  (0.2)  & 0.3  (0.1) & 0.0  (0.3) \\ 
C2 & - & - & -7.1  (0.0) & -4.2  (0.2) \\ \hline \hline
   \multicolumn{5}{c}{epoch 17 -- 23 (referenced feature: D8)}\\ \hline\hline
D1 & -6.8  (0.1) & -2.2  (0.1) & -1.5  (0.2) & -3.0  (0.1) \\
D2 & -7.2  (0.1) & -1.4  (0.3) & -1.9  (0.1) & -2.2  (0.3) \\
D3 & -7.0  (0.7) & -0.2  (0.5) & -1.7  (0.7) & -1.1  (0.5) \\
D4 & -6.7  (0.3) & -0.2  (1.3) & -1.4  (0.4) & -1.1  (1.3) \\
D5 & -4.3  (0.1) & 2.5  (0.1) & 1.0  (0.1) & 1.7  (0.1) \\
D6 & -1.8  (0.4) & 1.4  (0.4)  & 3.6  (0.5) & 0.5  (0.4) \\
D7 & -10.2  (0.5) & 7.5  (1.0) & -4.9  (0.5) & 6.6  (1.0) \\
D8 & - & - & 5.3  (0.1) & -0.9  (0.1)  \\ \hline
\multicolumn{5}{l}{All the parenthetic values indicate a relavant error.}\\
     \end{tabular} 
\end{table}

\begin{table*}[!htb]
\centering
  \caption{3D motions and inclinations}\label{tab:inclination}
   \begin{tabular}{ccccccc}\hline
ID & Region$^{a}$ & 2D Motions$^{b}$ & $V_{\rm LSR}$ - $V_{\rm sys}$$^{c}$ & $\Delta V_{\rm LSR}$ & Position Angle & Inclination$^{d}$ \\
    &            & (km s$^{-1}$) &  (km s$^{-1}$)  &  (km s$^{-1}$) & ($^{\circ}$) & ($^{\circ}$) \\ \hline
A1 & C & 9.6  (1.2) &-18.9  & 0.7  & -52.0 $^{+8.8}$$_{-11.0}$ &27.0  (3.0)  \\
A2 & E & 14.7  (2.8) &-38.8  & 0.4  & 62.8 $^{+8.4}$$_{-11.2}$ & 20.8  (3.6)  \\
A3 & C & 19.7  (1.1) &-47.7  & 2.4  & -131.2 $^{+4.6}$$_{-4.4}$& 22.4  (1.1)  \\
A4 & C & 17.7  (3.3) &-49.2  & 0.8  & -177.9 $^{+15.4}$$_{-14.8}$ &19.8 (3.5)  \\
A5 & E & 18.3  (0.4) &-59.7  & 0.8  & 82.7 $^{+7.1}$$_{-7.1}$& 17.1  (0.3)  \\
A6 & E & 29.9  (12.0) &-66.4  & 1.4  & -3.5 $^{+2.0}$$_{-4.6}$& 24.2  (9.2)  \\
A7 & E & 25.9  (0.6) &-68.8  & 0.7  & 25.9 $^{+0.9}$$_{-0.8}$& 20.7  (0.5)  \\
B1 & C & 11.6  (2.6) &-47.0  & 1.7  & 179.2 $^{+13.8}$$_{-14.1}$& 13.9  (3.0)  \\
B2 & C & 11.1  (3.0) &-48.8  & 1.3  & -52.1 $^{+20.3}$$_{-14.9}$ & 12.8  (3.3)  \\
B3 & C & 1.8  (1.8) &-51.7  & 1.4  & 110.7 $^{+212.7}$$_{-25.5}$ & 2.0  (2.0)  \\
B4 & E & 28.5  (1.0) &-66.9  & 0.6  & 61.2 $^{+2.1}$$_{-2.2}$& 23.1  (0.7)  \\
C1 & C & 53.2  (5.6) &-27.4  & 0.5  & -115.1 $^{+9.4}$$_{-9.9}$& 62.7  (2.7)  \\
C2 & C & 66.6  (0.8) &-39.5  & 0.6  & -120.6 $^{+1.2}$$_{-1.2}$& 59.3  (0.3)  \\
C3 & C & 31.9  (1.4) &-42.6  & 1.1  & -85.3 $^{+5.8}$$_{-5.5}$& 37.5  (1.3)  \\
C4 & C & 56.8  (2.7) &-47.0  & 2.0  & -88.8 $^{+6.4}$$_{-6.3}$& 50.4  (1.4)  \\
C5 & C & 31.2  (10.0) &-51.0  & 1.0  & -57.0 $^{+25.6}$$_{-30.2}$& 31.5  (8.8)  \\
C6 & C & 44.1  (8.7) &-52.6  & 1.4  & -103.6 $^{+28.3}$$_{-26.8}$& 40.0  (6.0)  \\
C7 & E & 2.0  (0.4) &-102.7  & 1.2  & 95.3 $^{+53.1}$$_{-59.7}$& 1.1  (0.2)  \\
D1 & W & 27.3  (0.9) &4.9  & 0.2  & -153.0 $^{+3.2}$$_{-3.1}$& 79.8  (0.3)  \\
D2 & W & 23.3  (2.0) &4.6  & 0.4  & -140.3 $^{+5.9}$$_{-5.3}$& 78.8  (1.0)  \\
D3 & C & 16.0  (5.0) &-48.9  & 1.3  & -122.6 $^{+18.2}$$_{-24.4}$& 18.1  (5.4)  \\
D4 & C & 14.1  (6.9) &-51.1  & 1.2  & -127.0 $^{+52.2}$$_{-29.4}$& 16.3  (7.4)  \\
D5 & C & 15.7  (0.7) &-53.8  & 0.7  & 32.0 $^{+3.4}$$_{-3.3}$& 16.3  (0.7)  \\
D6 & C & 29.0  (3.6) &-55.6  & 1.0  & 81.5 $^{+6.7}$$_{-8.2}$& 27.5  (3.0)  \\
D7 & C & 66.6  (6.9) &-59.1  & 1.0  & -36.4 $^{+6.5}$$_{-7.4}$& 48.4  (3.1)  \\
D8 & E & 43.4  (0.4) &-103.4  & 0.8  & 99.1 $^{+0.6}$$_{-0.6}$& 22.8  (0.2)   \\ \hline
\multicolumn{7}{l}{$^{a}$ "E": East side, "C": Central cluster, "W": West end}\\ 
\multicolumn{7}{l}{$^{b}$ Absolute values of internal proper motions at 1.7 kpc. }\\ 
\multicolumn{7}{l}{$^{c}$ $V_{\rm sys}$ of -5.0 km s$^{-1}$ was subtracted. }\\ 
\multicolumn{7}{l}{$^{d}$ Angles between 3D motions and the LOS. }\\ 
\multicolumn{7}{l}{All the parenthetic values indicate a relevant error.}\\ 
     \end{tabular} 
\end{table*}

\end{document}